\documentclass[%
prx,
 amsmath,amssymb,
 aps,
floatfix,
twocolumn,
longbibliography,
showkeys
]{revtex4-1}

\usepackage{graphicx}
\usepackage{dcolumn}
\usepackage{bm}
\usepackage{color}
\usepackage[mathlines]{lineno}
\usepackage[breaklinks=true,colorlinks=true,linkcolor=blue,urlcolor=blue,citecolor=blue]{hyperref}
\usepackage{ragged2e}
\usepackage{enumitem}
\usepackage{bibunits}
\usepackage{amsthm}
\usepackage{diagbox}

\renewcommand\thefigure{\arabic{figure}}

\begin{document}

\title[]{Small Bots, Big Impact: Solving the Conundrum of Cooperation in Optional Prisoner's Dilemma Game through Simple Strategies}

\author{Gopal Sharma$^1$}
\author{Hao Guo$^2$}
\author{Chen Shen$^3$}
\email{steven\_shen91@hotmail.com}
\author{Jun Tanimoto$^{1,3}$}
\email{tanimoto@cm.kyushu-u.ac.jp}

\affiliation{
\vspace{2mm}
\mbox{1. Interdisciplinary Graduate School of Engineering Sciences, Kyushu University, Fukuoka, 816-8580, Japan}
\mbox{2. School of Mechanical Engineering, and School of Artifcial Intelligence, Optics and ElectroNics (iOPEN),}
\mbox{Northwestern Polytechnical
University, Xi'an 710072, China}\\
\mbox{3. Faculty of Engineering Sciences, Kyushu University, Kasuga-koen, Kasuga-shi, Fukuoka 816-8580, Japan}
}

\date{\today}


\begin{abstract}
Cooperation plays a crucial role in both nature and human society, and the conundrum of cooperation attracts the attention from interdisciplinary research. In this study, we investigated the evolution of cooperation in optional prisoner's dilemma games by introducing simple bots. We focused on one-shot and anonymous games, where the bots could be programmed to always cooperate, always defect, never participate, or choose each action with equal probability. Our results show that cooperative bots facilitate the emergence of cooperation among ordinary players in both well-mixed populations and a regular lattice under weak imitation scenarios. Introducing loner bots has no impact on the emergence of cooperation in well-mixed populations, but it facilitates the dominance of cooperation in regular lattices under strong imitation scenarios. However, too many loner bots on a regular lattice inhibit the spread of cooperation and can eventually result in a breakdown of cooperation. Our findings emphasize the significance of bot design in promoting cooperation and offer useful insights for encouraging cooperation in real-world scenarios.
\end{abstract}

\maketitle

\section{Introduction}
Explaining the evolution of cooperation poses a scientific challenge because individuals who incur costs to benefit others may reduce their own fitness, which conflicts with the theory of "survival of the fittest"~\cite{darwin2004origin}. However, this prediction contrasts with the reality of ubiquitous altruistic behavior. Evolutionary game theory offers a powerful framework to address this issue~\cite{weibull1997evolutionary}. Within this framework, the prisoner's dilemma game, in particular, has been extensively used to study the evolution of cooperation in the context of social dilemma games, as it highlights the conflict between individual and collective interests~\cite{axelrod1980effective}. It has been shown that cooperation can survive through several reciprocity mechanisms, including group and kin selection~\cite{lehmann2007group,traulsen2006evolution}, direct and indirect reciprocity~\cite{axelrod1980effective,nowak2005evolution}, as well as network reciprocity~\cite{nowak1992evolutionary,szabo2007evolutionary}. Although these reciprocity mechanisms may appear materially different, they share a common principle, which is the mutual recognition of cooperators, also referred to as positive assortativity~\cite{pepper2002mechanism}. If the game is one-shot and anonymous where the above reciprocity mechanisms are fully excluded, the problem of cooperation can still be addressed by enabling players to punish wrongdoers~\cite{fehr2002altruistic}, reward altruists\cite{andreoni2003carrot,hilbe2010incentives}, or exercise their freedom to choose whether to participate in the game or not~\cite{hauert2002volunteering,szabo2002phase}. Several recent theoretical studies have shown that introducing committed cooperators can enable the survival of cooperation in well-mixed populations, even under unfavorable conditions such as one-shot and anonymous games without any reciprocity mechanisms~\cite{masuda2012evolution,nakajima2015evolutionary,cardillo2020critical}.

In the context of social dilemma games, committed cooperators refer to individuals who hold strong beliefs about altruistic behavior and do not change their behavior over time. While committed cooperators may not alter the advantage of defection over cooperation, their presence does increase the likelihood of interactions between cooperators. This ultimately leads to a greater propensity for cooperation in well-mixed situations when frequency-dependent updating rules are applied~\cite{masuda2012evolution,cardillo2020critical}. However, the cooperation-boosting effects of committed cooperators disappear when network reciprocity is considered~\cite{matsuzawa2016spatial}. Network reciprocity suggests that cooperators can survive by forming compact clusters to resist the invasion of defectors~\cite{nowak1992evolutionary}. The establishment of network reciprocity can be better understood by examining how cooperative clusters can form during their enduring period and how they can expand their territories during the expanding period~\cite{wang2013insight}. However, the presence of committed cooperators who consistently choose to cooperate can inhibit the formation of cooperative clusters and lower their spreading efficiency, resulting in a detrimental effect on cooperation rather than a beneficial one~\cite{matsuzawa2016spatial}.

Research has shown that committed minorities can play a crucial role in promoting cooperation, but in practical situations, such individuals are usually scarce. If a significant number of committed individuals are required to encourage cooperation, the impact of committed cooperators may be insignificant. With the advancement of artificial intelligence (AI), future populations will be more hybrid, consisting of both human and AI players. In the context of one-shot and anonymous games, where there are no reputation effects or repeated interactions, and where information is limited, bots can be designed using straightforward algorithms based on the concept of committed individuals~\cite{shirado2020interdisciplinary}. These bots are programmed to consistently choose one action and never change it over time. Although their strategy is simple, they can help solve complex problems that human players may find challenging. For example, they can aid in resolving cooperation problems~\cite{santos2019evolution,shirado2020interdisciplinary,shirado2020network}, tackling the collective risk dilemma~\cite{fernandez2022delegation,terrucha2022art}, and addressing the conundrum of punishment~\cite{shen2022simple}, and others~\cite{shirado2017locally}.

In this paper, we investigate the evolution of cooperation among human players in human-machine interactions by examining the impact of four types of bots on an optional prisoner's dilemma game. The game involves three actions: cooperation to benefit others ($C$), defection for self-interest ($D$), and loner for risk aversion ($L$). The four types of pre-designed bots are: (i) always choosing to cooperate, (ii) always choosing to defect, (iii) always choosing loner, and (iv) choosing cooperation, defection, and loner with equal probability. Additionally, we consider two representative population structures: a well-mixed population where players have an equal probability of interacting with others, and a regular lattice where players can only interact with their nearest neighbors. Our findings suggest that introducing cooperative bots can address the conundrum of cooperation under weak imitation strength scenarios in both well-mixed populations and regular lattices. Additionally, although introducing loner bots does not have any impact on the emergence of cooperation in well-mixed populations, it facilitates the dominance of cooperation under strong imitation scenarios in regular lattices. However, too many loner bots on a regular lattice inhibit the spread of cooperation and can eventually result in a breakdown of cooperation. Our results, taken together, indicate that the level of cooperation can be maximized by managing the behavior of simple bots.

\section{Model}
Our model involves four essential components, which are: (a) the payoff matrix, (b) population settings, (c) game dynamics, and (d) simulation settings. Each of these components is described in detail below.
 \subsection{Optional prisoner's dilemma game}
In the optional prisoner's dilemma game, players choose between cooperation, defection, or acting as a loner simultaneously. Cooperators incur a cost of $c$ but bring a benefit of $b$ to their opponent, while defectors do nothing. Loners choose to quit the game and receive a small, but positive payoff of $\sigma$, generating the same payoff for their opponent. If two cooperators meet, they both receive a payoff of $R=b-c$. Two defectors receive a payoff of $P=0$, while a cooperator meeting a defector incurs a cooperation cost of $S=-c$, and the defector receives a benefit of $T=b$. By rescaling the payoffs using $r=c/(b-c)$ and substituting, the payoffs become $T=1+r$, $R=1$, $P=0$, and $S=-r$. The resulting payoff matrix is shown in Table \ref{t01}. We use the concept of universal dilemma strength~\cite{wang2015universal} to quantify the dilemma extent, which is confirmed to be ${D_{g}}'={D_{r}}'= r$~\cite{tanimoto2021sociophysics}. This particular prisoner's dilemma game exhibits characteristics of both chicken-type dilemmas (arising from greed) and stag-hunt-type dilemmas (arising from fear).

\begin{table}
\caption{\label{t01} Payoff matrix for the optional prisoner's dilemma game.}
\begin{ruledtabular}
\begin{tabular}{cccc}
~   & $C$ & $D$  & $L$  \\
\hline
$C$ & $1$ & $-r$ & $\sigma$\\
$D$ & $1+r$ & $0$ & $\sigma$\\
$L$ & $\sigma$ & $\sigma$ & $\sigma$
\end{tabular}
\end{ruledtabular}
\end{table}

\begin{figure*}[htbp]
    \centering
\includegraphics[width=0.91\linewidth]{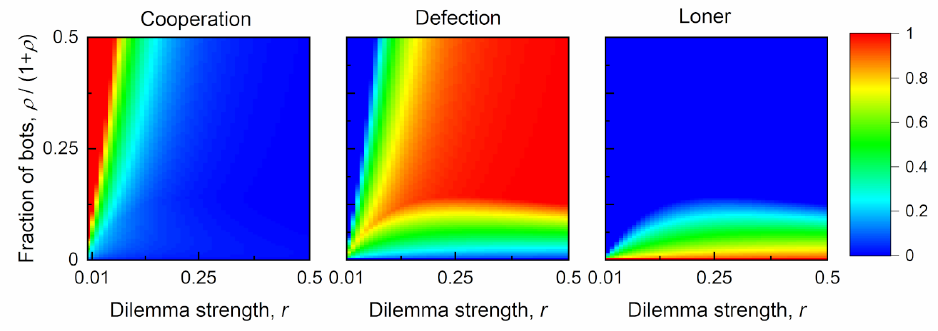}
   \caption{\textbf{In well-mixed populations, the introduction of cooperative bots can facilitate the emergence of cooperation among ordinary players, whereas in the absence of bots, loners tend to dominate the entire population.} Shown are the fraction of cooperation (left), defection (middle), and loner (right) at steady state, as a function of dilemma strength and density of bots in well-mixed situations. Here, bots were programmed to consistently choose cooperation and not change their action over time. The imitation strength was fixed at $\kappa^{-1}=10$.}
    \label{fig1}
\end{figure*}

\subsection{Population settings}
There are typically two population structures considered in research: well-mixed and structured. In a well-mixed population, individuals can interact with any other individual. We denote the fractions of cooperators, defectors, and loners as $x,y$, and $z$, respectively, such that $x+y+z=1$. We assume that the fraction of bots is $\rho$, and therefore the total population density is $1+\rho$. In contrast, a structured population only allows for local interactions, where individuals can only interact with their immediate neighbors. We represent the structured population as a two-dimensional regular lattice with periodic conditions and a degree of four, indicating that each node has four closest neighbors in the up, down, left, and right directions. The regular lattice has a size of $L \times L$. The entire population is divided into two types at random in a regular lattice, with $\rho$ proportion as bot players and the remaining proportion as normal players. In both well-mixed and regular lattice populations, normal players are presented with three options: cooperation, defection, and loner, each with an equal likelihood. Bot players are pre-programmed with four strategies: (i) always choosing cooperation, (ii) always choosing defection, (iii) always choosing loner, and (iv) choosing cooperation, defection, and loner with equal probability.
\subsection{Game dynamics}
The mean-field theory was utilized to analyze the human-machine game within an infinite and well-mixed population. We provide an example of the game dynamics for the cooperative bots, while additional details regarding the other three types of bots can be found in the Appendix. As described in the population settings section, the expected payoff for each actor can be represented by:

\begin{equation}
\begin{array}{l}
\Pi_{C}  = \frac{x-yr+z\sigma+\rho}{1+\rho} \\
\Pi_D  = \frac{(1+r)(x+\rho)+\sigma z}{1+\rho} \\
\Pi_{L}  = \sigma\\
\end{array}.
\label{eq01}
\end{equation}
We adopt the pairwise comparison rule, where the imitation probability depends on the payoff difference between two randomly selected players. If the randomly selected players choose the same action, nothing happens. Otherwise, the probability that action $i$ replaces action $j$ is:
\begin{equation}
\begin{array}{l}
P_{j \to i}  = 1/(1+e^{(\Pi_{j}-\Pi_{i})/\kappa}) \\
\end{array},
\label{eq02}
\end{equation}
where $i\not=j \in \{C,D,L\}$, and $\kappa^{-1}$ is the imitation strength such that $\kappa^{-1}>0$, and we fixed the value of $\kappa^{-1}$ at 10 throughout the paper. The evolutionary dynamics of the well-mixed and infinite population under the imitation rule are represented by:
\begin{equation}
\begin{array}{l}
\dot{x} = \frac{2}{1+\rho}((x+\rho)yP_{D \to C} + (x+\rho)zP_{L \to C}-xyP_{C \to D}\\-xzP_{C \to L})\\
\dot{y} = \frac{2}{1+\rho}(xyP_{C \to D} + yzP_{L \to D}-(x+\rho)yP_{D \to C}\\-yzP_{D \to L})\\
\dot{z} = \frac{2}{1+\rho}(xzP_{C \to L} +yzP_{D \to L}-(x+\rho)zP_{L \to C}\\-yzP_{L \to D})\\
\end{array}.
\label{eq03}
\end{equation}

To generate results for networked populations, we employed Monte Carlo simulations. These simulations involved the following steps: First, each node was designated as a bot with probability $\rho$, while the probability that each node was designated as a normal player was $1-\rho$. For the population of normal players, each player was randomly assigned as a cooperator ($C$), defector ($D$), or loner ($L)$. Normal players and bots residing on networks received their payoffs by interacting with all direct neighbors. In each round of the game, a randomly selected normal player $i$ updated its action by imitating the action of a randomly selected neighbor $j$, which could be either a bot or a normal player. The probability of imitation was determined by the pairwise Fermi function (PW-Fermi), as defined in Eq. \ref{eq02}.

\begin{figure*}[htbp]
    \centering
\includegraphics[width=0.91\linewidth]{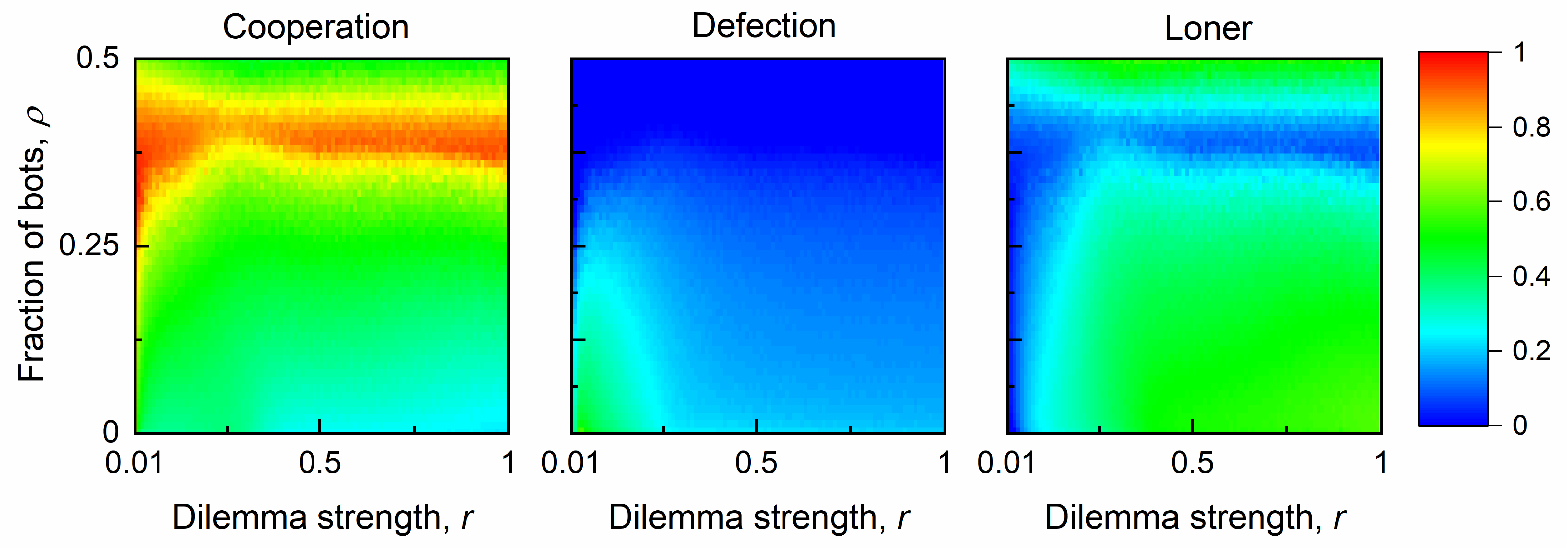}
   \caption{\textbf{In a regular lattice, the introduction of loner bots can facilitate the emergence of cooperation among ordinary players, whereas too dense of loner bots results in a breakdown of cooperation.} Shown are the fraction of cooperators, defectors, and loners among ordinary players as a function of dilemma strength and density of bots in a regular lattice. Here bots were programmed to consistently choose loner action and not change their action over time. The imitation strength was fixed at $\kappa^{-1}=10$.}
    \label{fig2}
\end{figure*}

\subsection{Simulation settings}
The game dynamics for a single Monte Carlo step were executed $L*L$ times, allowing each player to modify their action at least once on average. We set the value of $L$ at 100 for the entire study. In order to achieve an equilibrium state, all simulations were typically conducted for 50,000 steps. This ensured that there was no obvious increasing or decreasing trend for the fractions of each actor. The proportion of each action was then calculated by averaging the results of the last 5000 steps, providing a more stable and accurate representation of the equilibrium state. To reduce the impact of initial conditions, we calculated the data by averaging the results of 20 independent runs. Our focus is solely on the impact of bots on the evolution of cooperation among normal players. Therefore, when calculating the fractions of each action in the population, we disregard the fraction of bot players.

\section{Results}
\subsection{Well-mixed populations}
Loners receive a small, but positive payoff by opting out and generate the same payoff for their opponent. This definition allows for the coexistence of cooperators, defectors, and loners through the effect of cyclic dominance in well-mixed populations~\cite{hauert2002volunteering}. However, the cooperation-sustaining effect by loners is only possible in the public goods game, and not in the prisoner's dilemma game~\cite{szabo2002evolutionary}. In fact, the prisoner's dilemma game can lead to a pure loner state in well-mixed populations. Although the cyclic dominance pattern can still exist along the edges (i.e., $C$ giving way to $D$, $D$ giving way to $L$, who gives way to $C$), the invasion direction from $L$ to $C$ requires a certain level of cooperation~\cite{mathew2009does}. The addition of simple bots to the optional prisoner's dilemma game can significantly alter the fate of cooperators. Rather than being entirely eliminated by defectors, cooperators can now thrive and even establish dominance over the entire population with the assistance of cooperative bots (Figure.\ref{fig1}).

Introducing cooperative bots doesn't change the payoff scheme for human players, but it does increase the chances of human players encountering other cooperators. This higher likelihood of encountering cooperators helps to reduce the advantage that loners have over cooperators and defectors. However, it also creates an opportunity for free-riding defectors to exploit the system. If human players place less emphasis on material gains when updating their actions (in a weak imitation scenario), natural selection is likely to favor the dominance of cooperation (see Figure \ref{figa1}). Conversely, in a strong imitation scenario, where players tend to copy the behavior of those who are most successful, the entire population would be dominated by defectors (see Figure \ref{figa1}). The introduction of defective or loner bots cannot mitigate the advantage that loners hold over cooperators and defectors in the absence of bots. Consequently, these two types of bots do not affect the emergence of cooperation, as illustrated in the top and middle rows of Figure \ref{figa2}. If the actions of bot players were programmed to choose cooperation, defection, and loner with equal probability, then the diversity of bot actions reduces the cooperation-promoting effect by cooperative bots (refer to the bottom row of Figure \ref{figa2}).

While the pure loner state of the optional prisoner's dilemma game is observed in well-mixed populations, regular lattices facilitate the stable coexistence of cooperators, defectors, and loners through cyclic dominance, in which cooperation give ways to defection, who give ways to loner, who in turn give ways to cooperation~\cite{szabo2002evolutionary}. This difference prompts further investigation into how the addition of simple bots influences the evolution of cooperation in structured populations.

\begin{figure}[htbp]
    \centering
\includegraphics[width=0.98\linewidth]{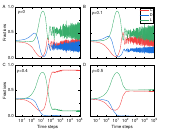}
   \caption{\textbf{The introduction of loner bots can disrupt the cyclic dominance that typically sustains cooperation among cooperators, defectors, and loners in a regular lattice. However, adding loner bots can improve the level of cooperation among human players by reducing the number of defectors. } Shown are the fractions of cooperators, defectors, and loners as a function of simulation steps under different values of density of bots. The dilemma strength was fixed at $r=0.5$, while the values of $\rho$ were set to 0, 0.1, 0.4, and 0.5 for panels A to D.}
    \label{fig3}
\end{figure}

\subsection{Structured populations}
Consistent with the cooperation-promoting effect by cooperative bots in well-mixed populations, introducing cooperative bots to the optional prisoner's dilemma game on a regular
lattice still promotes cooperation among human players under the weak imitation scenarios (Figure.\ref{figa03}). However, the required imitation strength threshold to achieve cooperation prosperity decreases. For example, introducing cooperative bots can decrease cooperation when the imitation strength equals $\kappa=10^{-1}$ (as shown in the top row of Figure \ref{figa3}), whereas the same imitation strength value ensures increased cooperation in well-mixed populations. A similar cooperation-decreasing phenomenon can be also observed when using defective bots or when bots randomly choose among cooperation, defection, and being a loner (as shown in the middle and bottom rows of Figure \ref{figa3}). In contrast with the results on well-mixed population in which introducing loner bots does not have any impact on the emergence of cooperation, the cooperation-promoting phenomenon can be observed if bots are programmed to always act as loners and not change their behavior over time, as demonstrated in Figure \ref{fig2}. This figure shows the fraction of cooperation (left panel), defection (middle panel), and loner (right panel) as a function of dilemma strength and the density of loner bots, obtained on a regular lattice using Monte Carlo simulations. Interestingly, introducing loner bots actually increases the level of cooperation over the entire range of dilemma strength. Moreover, there exists an optimal density of loner bots at which cooperation is maximized, and further increasing the density of loner bots results in the breakdown of cooperation. In addition, the cooperation-promoting effect by loner bots is restricted in strong imitation scenario, while loners tend to dominant the population in the weak imitation scenario (Figure.\ref{figa4}). To reveal the potential mechanism behind the effect of loner bots on cooperation, we examine the overall evolution of actor abundance along the temporal dimension, as well as the spatial and temporal dimensions by analyzing the time-series of each actor and the evolutionary snapshots.

The dynamics of the three actors and their interactions can be seen in Figure \ref{fig3}, where the fractions of cooperators, defectors, and loners are plotted as a function of Monte Carlo steps for different values of $\rho$: 0, 0.1, 0.4, and 0.5. This figure illustrates how the behavior of one actor affects the others, and provides insight into the system's overall evolution. When $\rho=0$, our model reduces to the traditional optional prisoner's dilemma game where cooperation coexists with defection and loner through cyclic dominance. As shown in Figure \ref{fig3}A, defectors initially reach their peak and then give way to loners, who in turn give way to cooperators. This process repeats until the system reaches its stable state, where the abundances of each actor fluctuate around their average value. Similar cyclic dominance occurs for a small density of loner bots (i.e., $\rho=0.1$), but the stationary defection level decreases (Figure \ref{fig3}B). However, for a large density of bots (i.e., $\rho=0.4$ and $\rho=0.5$), the cyclic dominance phenomenon disappears, and loners initially reach their peak followed by cooperators, while defectors decrease and go extinct directly (Figure.\ref{fig3}C and Figure.\ref{fig3}D). The evolutionary dynamics at $\rho=0.4$ and $\rho=0.5$ are similar, but the survival periods of defection in the former situation are longer than those in the latter case.

\begin{figure*}[htbp]
    \centering
\includegraphics[width=0.96\linewidth]{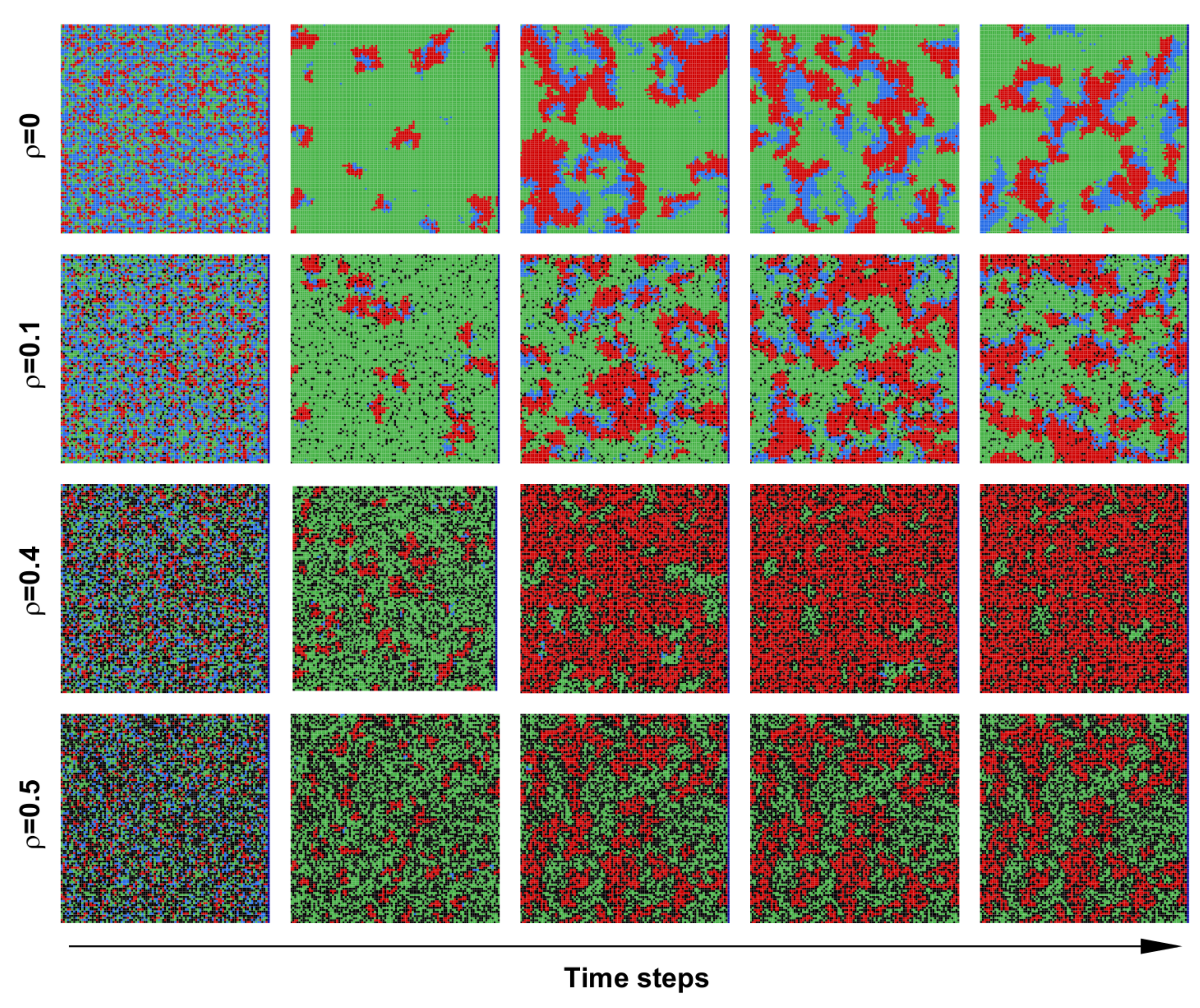}
   \caption{\textbf{Evolutionary snapshots reveal that a high density of solitary bots can create a barrier for the spread of cooperation, leading to the breakdown of cooperative behavior.} Shown are the evolutionary snapshots taken at different values of bot intensity, with the dilemma strength fixed at $r=0.5$ and the value of $\rho$ fixed at 0, 0.1, 0.4, and 0.5 from top to bottom rows. The snapshots in the first column were taken at the initial step, where actions were distributed randomly for both human players and bot players. The snapshots in the remaining columns were taken at different time steps, making the figure as illustrative as possible.}
    \label{fig4}
\end{figure*}

Figure \ref{fig4} shows the evolutionary snapshots under the same parameter values as Figure \ref{fig3}. In this figure, cooperators, defectors, and loners are represented by red, blue, and green, respectively. Bot players are represented by dark green. When $\rho=0$, our model is reduced to the traditional optional prisoner's dilemma game, and we observe the expected traveling waves of cooperators, followed by defectors, then loners, and finally cooperators again (top row of Figure \ref{fig4}). Introducing a small density of loner bots does not significantly affect the traveling waves of cooperators but reduces the chasing ability of defectors, as shown in the second row of Figure \ref{fig4}. Here, loner bots tend to be surrounded by compact cooperative clusters, and defective clusters are much smaller than those observed in the scenario without bots. However, when the density of loner bots is high (i.e., $\rho=0.4$), defectors struggle to survive the invasion of loners (see the second column and third row of Figure \ref{fig4}) and gradually go extinct. Without the threat of defectors, cooperators take over the territory of loners until some lucky loners are surrounded by loner bots. If the density of loner bots is too large (i.e., $\rho=0.5$), although defectors can still be eliminated by loners, cooperation cannot spread further in the population since cooperative clusters can easily be fully surrounded by loner bots (bottom row in Figure \ref{fig4}), leading to a decreased cooperation level compared to the situation with $\rho=0.4$.


\section{Discussion}
To discuss, in this paper, we investigated the impact of simple bots on the evolution of cooperation in the optional prisoner's dilemma game, using four types of bots and two representative populations. Despite their simplicity and lack of intelligent algorithms, these bots play a crucial role in determining the evolution of cooperation. Our findings demonstrate that the introduction of cooperative bots can enhance the emergence of cooperation among ordinary players in both well-mixed populations and regular lattices under weak imitation scenarios. However, regular lattices require a weaker imitation strength scenario to achieve the same level of cooperation. In contrast to the results on well-mixed populations, where adding loner bots does not impact the emergence of cooperation, we observed that introducing loner bots promotes cooperation on a regular lattice under a strong imitation scenario. Specifically, we found that an optimal density of loner bots exists, at which the highest level of cooperation is achieved; otherwise, cooperation levels decrease. Introducing defective bots does not significantly impact the evolution of cooperation in well-mixed populations, and diversity in bot actions reduces the cooperation-promoting effect by cooperative bots. Similar cooperation-devastating results can be observed for defective bots or considering the diversity of bot actions. Our results, taken together, indicate that the level of cooperation can be maximized by managing the behavior of simple bots.

Voluntary participation in public good games can be an effective way to promote cooperation in well-mixed populations, even in the absence of other reciprocity mechanisms like reputation, iterated interactions, or group selection. However, it's important to note that cooperation is only likely to emerge if the group size is larger than two; otherwise, pairwise interactions lead to the homogeneous loner state~\cite{hauert2002replicator}.  While we focused on the simplest form of the game with only two players and did not explore more complex multiplayer scenarios in this paper, the results on regular lattices may still hold as the cyclic dominance effect is valid for both pairwise and multiplayer games~\cite{szabo2002phase}. It's worth noting that the results on well-mixed populations may differ significantly as periodic oscillations can only be observed for multiplayer games~\cite{hauert2002replicator}. Therefore, future research should investigate how the existence of simple bots affects period oscillations and which types of bots can promote cooperation in well-mixed populations, particularly in the context of public good games.

The critical assumptions in our human-machine framework are the one-shot game and anonymous interaction. It's important to clarify that a one-shot game differs from a strict one-round setting game. In a one-round game, the game can occur only once for all players, but it can still be played multiple times in a one-shot game setting as long as the opponents are strangers who have never met before. Unlike a strict one-round game, where players cannot revise their actions, players in a one-shot game can revise their actions based on their past experiences. This feature of the one-shot setting makes it possible for bot players to improve the cooperation level among human players, as experimental evidence shows that cooperative behavior can cascade among human players in social networks~\cite{fowler2010cooperative}. The existence of bots that always cooperate and do not change their actions over time may increase the probability that human players pass their obtained help to new strangers. This is possible under the weak imitation scenario, in which human players may still pass on this help even if they know they may endure some losses. However, our results are not limited to the weak imitation scenario since loner bots can realize high cooperation levels among human players in structured populations under the strong imitation scenario. These results indicate that cooperation can be improved through effective bot design management by considering specific population structures or the features of human learning moods.

The second anonymous interaction assumption assumes that there are no reciprocity mechanisms such as indirect reciprocity, direct reciprocity, reputation, or costly signals~\cite{gintis2001costly,axelrod1980effective,nowak2005evolution,bolton2005cooperation,cooper1996cooperation}. This assumption implies that both human and bot players cannot use information about their opponents in their decision-making process. Moreover, human players are not aware of the existence of bot players. These assumptions guarantee that human players update their actions similarly when they play against other human players or bot players. However, these assumptions also limit the intelligence of bot strategies since they cannot access information about their opponent. Although the past experiences of bot players are available in our framework, we intentionally disregarded them when designing the bots' strategies. Our main focus was to investigate the impact of simple bots on cooperation, and considering past experiences of bots would have added unnecessary complexity to the analysis. However, in future studies, the past experiences of bots could be incorporated using reinforcement learning methods to develop more sophisticated bot strategies~\cite{tampuu2017multiagent,bucsoniu2010multi}. By doing so, we can obtain a more realistic and nuanced understanding of the role of bots in promoting cooperation.

While our assumptions may not hold in more complex and realistic game situations where human-machine interactions are not one-shot but repeated, or where the reputation of opponents is also available, our results provide useful insights into the emergence of cooperation. Nonetheless, to address these more complex scenarios, it is necessary to develop more intelligent machines. Our findings can serve as a valuable starting point for future studies in this direction.

\section*{Article information}
\paragraph*{Acknowledgements.}
We acknowledge support from (i) a JSPS Postdoctoral Fellowship Program for Foreign Researchers (grant no. P21374), and an accompanying Grant-in-Aid for Scientific Research from JSPS KAKENHI (grant no. JP 22F31374), and the National Natural Science Foundation of China (grant no.~11931015) to C.\,S. as a co-investigator, (ii) MITSUI2207 from JEES awarded to G.\,S as a Scholarship (iii) the grant-in-Aid for Scientific Research from JSPS, Japan, KAKENHI (grant No. JP 20H02314) awarded to J.\,T.
\paragraph*{Author contributions.}
C.\,S. and J.\,T. conceived research. G.\,S., H.\,G, and C.\,S. performed simulations. All co-authors discussed the results and wrote the manuscript.
\paragraph*{Conflict of interest.} Authors declare no conflict of interest.

\vspace{5mm}

\appendix{{\bf\Large Appendix}}

\setcounter{equation}{0}
\setcounter{figure}{0}
\renewcommand\theequation{A\arabic{equation}}
\renewcommand\thefigure{A\arabic{figure}}


\begin{figure}[htbp]
    \centering
\includegraphics[width=0.91\linewidth]{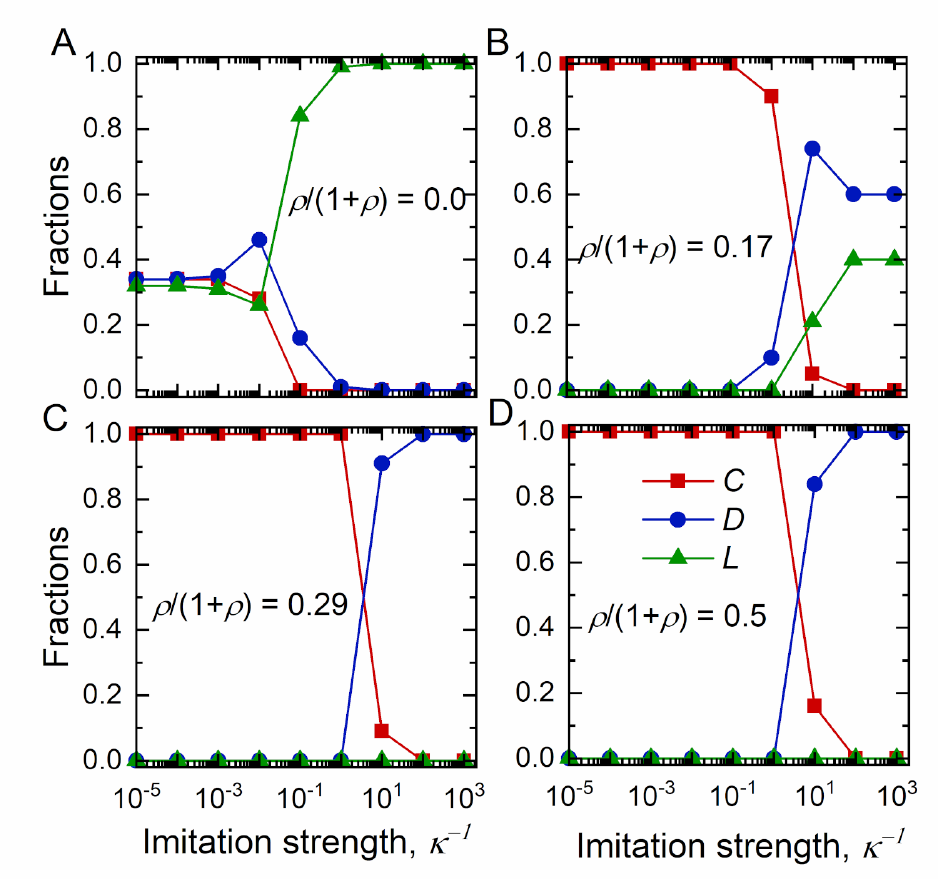}
   \caption{\textbf{The impact of cooperative bots on promoting cooperation in well-mixed populations depends on the strength of imitation. In weak imitation scenarios, cooperative bots can effectively increase cooperation rates, but in strong imitation scenarios, defection tends to dominate.} Shown are the proportions of cooperation, defection, and loners relative to imitation strength $\kappa^{-1}$, with fixed parameters of $r=0.2$, and $\rho$ values of 0 (panel A), 0.2 (panel B), 0.6 (panel C), and 1 (panel D).}
    \label{figa1}
\end{figure}

\begin{figure}[htbp]
    \centering
\includegraphics[width=0.96\linewidth]{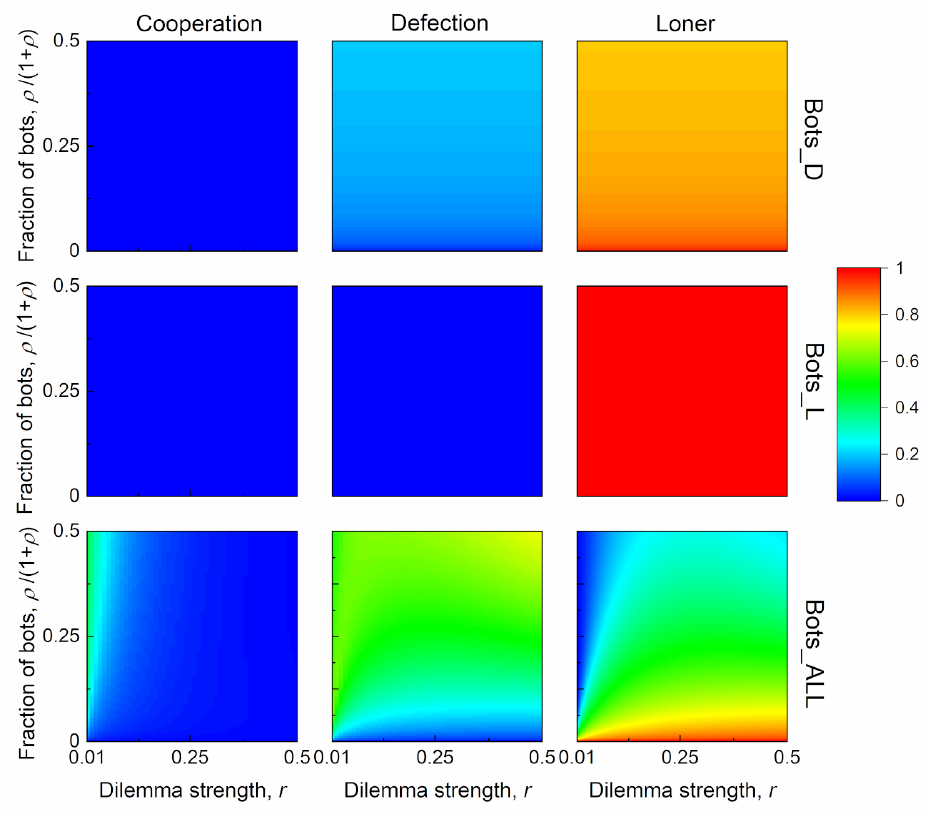}
   \caption{\textbf{Introducing defective or loner bots into the optional prisoner's dilemma game does not have any effect on the emergence of cooperative behavior in well-mixed populations, and the diversity of bot actions reduces the promotion effect of cooperation by cooperative bots.} The three panels from top to bottom correspond to different bot behaviors: (i) always defect (marked as $Bots\_D$), (ii) always loner (marked as $Bots\_L$), and (iii) random choice among cooperation, defection, and loner with equal probability (marked as $Bots\_ALL$). The imitation strength was fixed at $\kappa^{-1}=10$.}
    \label{figa2}
\end{figure}

\begin{figure}[htbp]
    \centering
\includegraphics[width=0.91\linewidth]{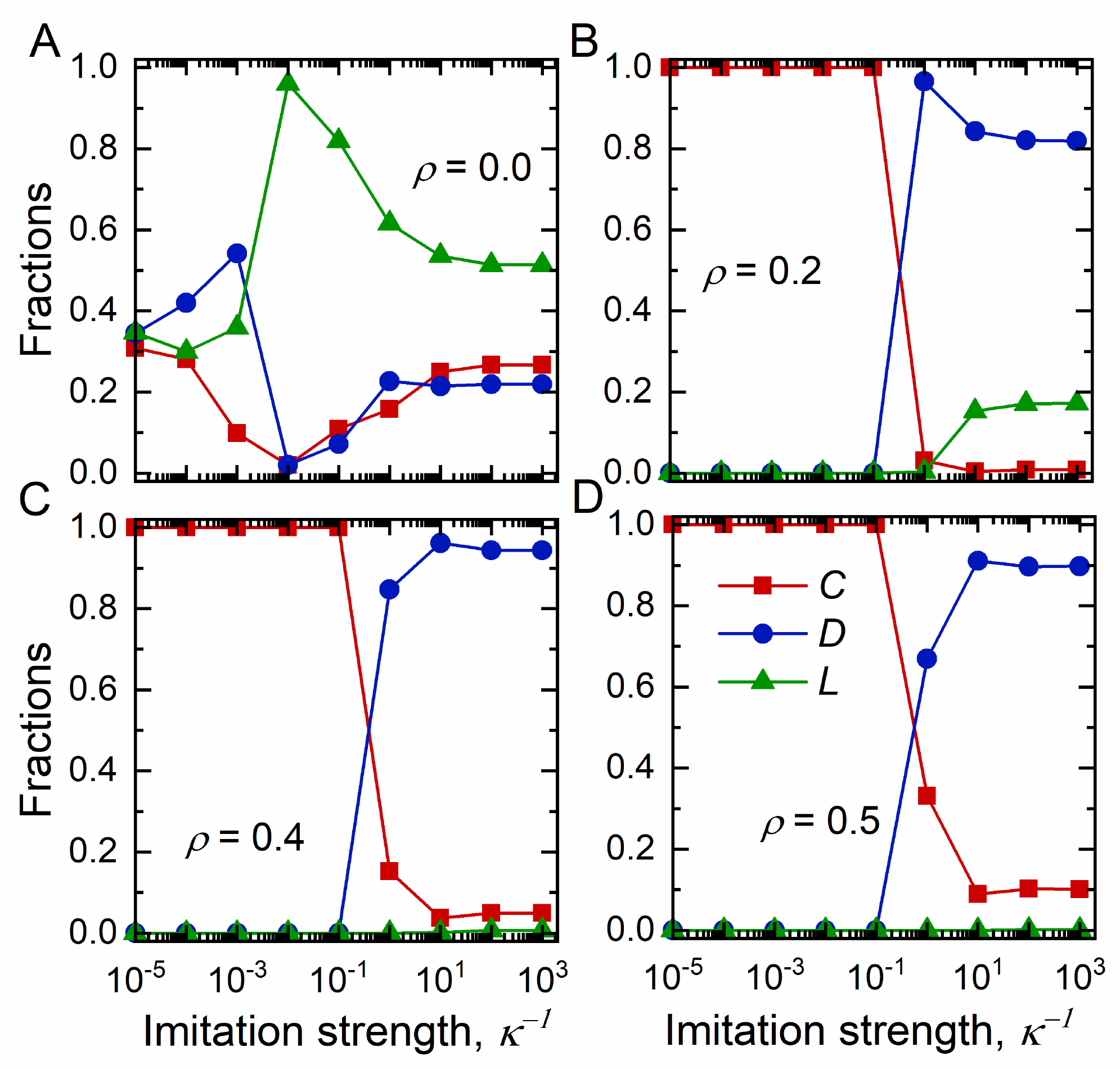}
   \caption{\textbf{The cooperation-promoting effect by cooperative bots in regular lattice still holds under weak imitation strength scenarios, while defectors tend to dominate the population under the strong imitation scenarios.} Shown are the proportions of cooperation, defection, and loners relative to imitation strength $\kappa^{-1}$, with fixed parameters of $r=0.5$, and $\rho$ values of 0 (panel A), 0.2 (panel B), 0.4 (panel C), and 0.5 (panel D). }
    \label{figa03}
\end{figure}

\begin{figure}[htbp]
    \centering
\includegraphics[width=0.96\linewidth]{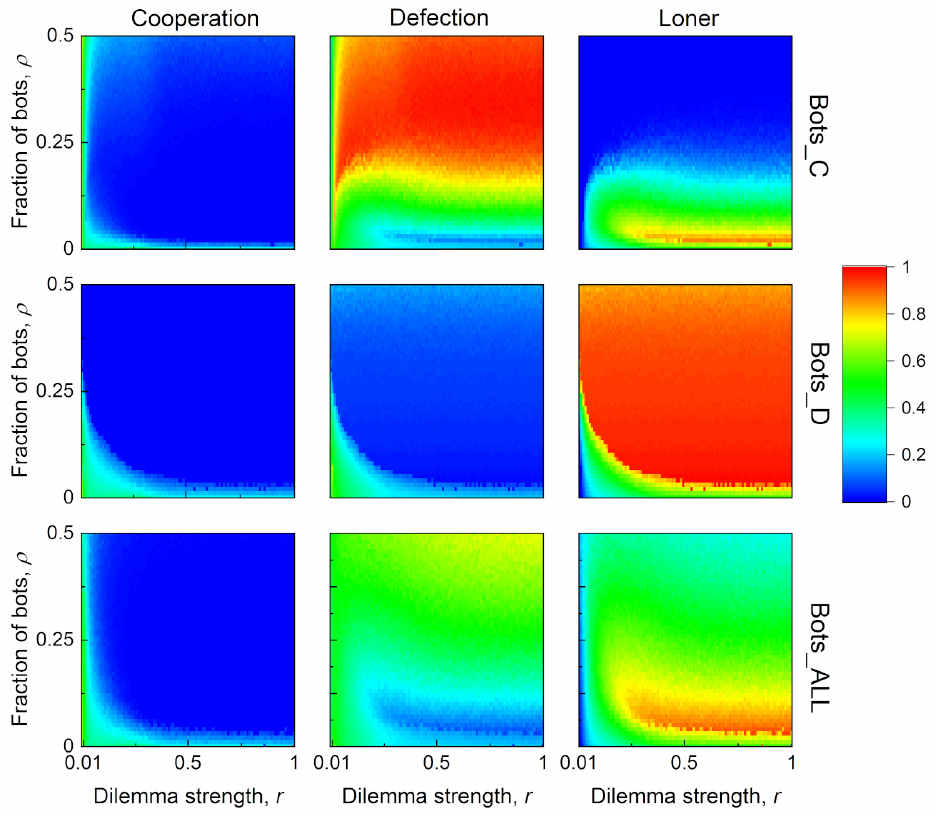}
   \caption{\textbf{Introducing cooperative or defective bots into the optional prisoner's dilemma game bring little impact on the emergence of cooperation in a regular lattice, and the diversity of bot actions reduces the promotion effect of cooperation by loner bots.} The three panels from top to bottom correspond to different bot behaviors: (i) always defect (marked as $Bots\_D$), (ii) always loner (marked as $Bots\_L$), and (iii) random choice among cooperation, defection, and loner with equal probability (marked as $Bots\_ALL$). The imitation strength was fixed at $\kappa^{-1}=10$.}
    \label{figa3}
\end{figure}

\begin{figure}[htbp]
    \centering
\includegraphics[width=0.91\linewidth]{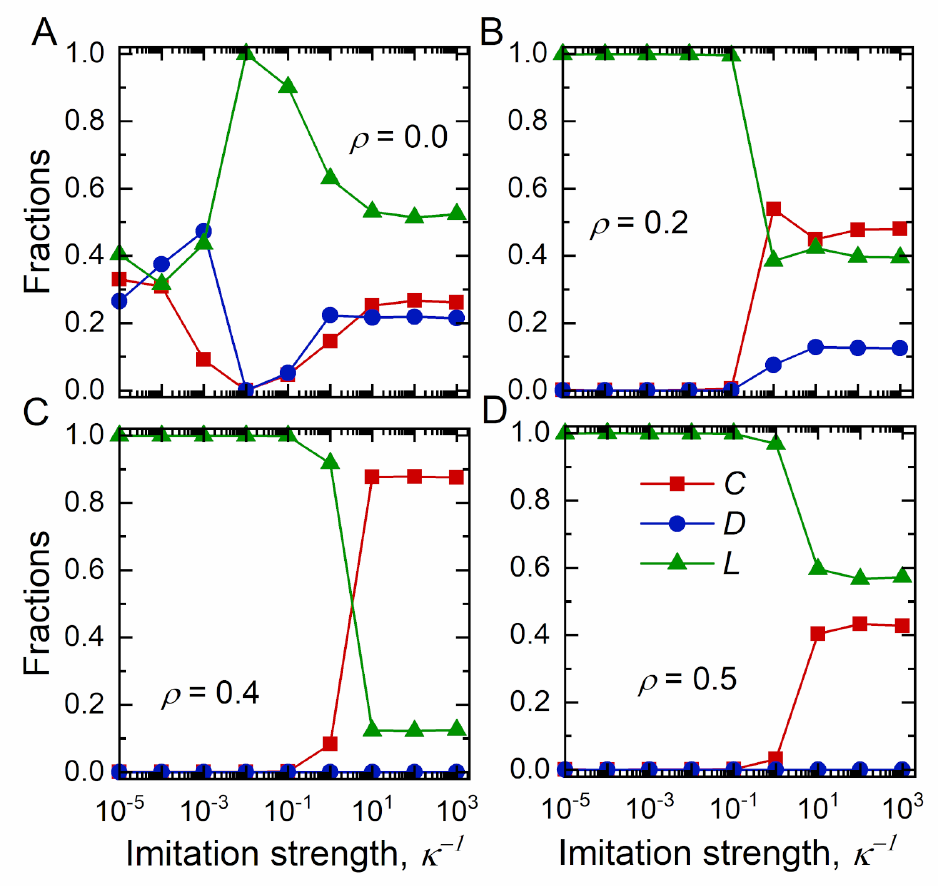}
   \caption{\textbf{The impact of loner bots on promoting cooperation in a regular lattice depends on the strength of imitation. In scenarios with weak imitation, loners tend to dominate the population, whereas in scenarios with strong imitation, loner bots can effectively increase cooperation rates.} Shown are the proportions of cooperation, defection, and loners relative to imitation strength $\kappa^{-1}$, with fixed parameters of $r=0.5$, and $\rho$ values of 0 (panel A), 0.2 (panel B), 0.4 (panel C), and 0.5 (panel D). }
    \label{figa4}
\end{figure}
\section*{Game Dynamics for Scenarios with Defective, Loner, and Randomized Bots}
\noindent\textbf{Remark 1: Defective bots.} Let $x,y$, and $z$ denote the fractions of cooperators ($C$), defectors ($D$), and loners ($L$) in a well-mixed and infinite population, such that $x+y+z=1$. We assume the fraction of bots is $\rho$, and the bots are designed always to choose $D$ and never change their action. The total population density is then $1+\rho$, and the expected payoffs of each of the actors are as follows:
\begin{equation}
\begin{array}{l}
\Pi_{C}  = \frac{x-(y+\rho)r+z\sigma}{1+\rho} \\
\Pi_D  = \frac{(1+r)x+z\sigma}{1+\rho} \\
\Pi_{L}  = \sigma\\
\end{array}.
\label{eqs01}
\end{equation}
We adopt the pairwise comparison rule, where the imitation probability depends on the payoff difference between two randomly selected players. If the randomly selected players choose the same action, nothing happens. Otherwise, the probability that action $i$ replaces action $j$ is:
\begin{equation}
\begin{array}{l}
P_{j \to i}  = 1/(1+e^{(\Pi_{j}-\Pi_{i})/\kappa}) \\
\end{array}.
\label{eqs02}
\end{equation}
Where $i\not=j \in \{C,D,L\}$, and $\kappa^{-1}$ is the imitation strength such that $\kappa^{-1}>0$. The evolutionary dynamics of the well-mixed and infinite population under the imitation rule are represented by:
\begin{footnotesize}
\begin{equation}
\begin{array}{l}
\dot{x} = \frac{2}{1+\rho}(xyP_{D \to C} + xzP_{L \to C}-x(y+\rho)P_{C \to D}-xzP_{C \to L})\\
\dot{y} = \frac{2}{1+\rho}(x(y+\rho)P_{C \to D} + (y+\rho)zP_{L \to D}-xyP_{D \to C}-yzP_{D \to L})\\
\dot{z} =\frac{2}{1+\rho}(xzP_{C \to L} +yzP_{D \to L}-xzP_{L \to C}-(y+\rho)zP_{L \to D})\\
\end{array}.
\label{eqs03}
\end{equation}
\end{footnotesize}

\noindent\textbf{Remark 2: loner bots} Let $x,y$, and $z$ denote the fractions of cooperators ($C$), defectors ($D$), and loners ($L$) in a well-mixed and infinite population, such that $x+y+z=1$. We assume the fraction of bots is $\rho$, and the bots are designed always to choose $L$ and never change their action. The total population density is then $1+\rho$, and the expected payoffs of each of the actors are as follows:
\begin{equation}
\begin{array}{l}
\Pi_{C}  = \frac{x-yr+(z+\rho)\sigma}{1+\rho} \\
\Pi_D  = \frac{(1+r)x+(z+\rho)\sigma}{1+\rho} \\
\Pi_{L}  = \sigma\\
\end{array}.
\label{eqs04}
\end{equation}
We adopt the pairwise comparison rule, where the imitation probability depends on the payoff difference between two randomly selected players. If the randomly selected players choose the same action, nothing happens. Otherwise, the probability that action $i$ replaces action $j$ is:
\begin{equation}
\begin{array}{l}
P_{j \to i}  = 1/(1+e^{(\Pi_{j}-\Pi_{i})/\kappa}) \\
\end{array}.
\label{eqs05}
\end{equation}
Where $i\not=j \in \{C,D,L\}$, and $\kappa^{-1}$ is the imitation strength such that $\kappa^{-1}>0$. The evolutionary dynamics of the well-mixed and infinite population under the imitation rule are represented by:
\begin{footnotesize}
\begin{equation}
\begin{array}{l}
\dot{x} = \frac{2}{1+\rho}(xyP_{D \to C} +xzP_{L \to C}
-xyP_{C \to D}-x(z+\rho)P_{C \to L})\\
\dot{y} = \frac{2}{1+\rho}(xyP_{C \to D} + yzP_{L \to D}-xyP_{D \to C}-y(z+\rho)P_{D \to L})\\
\dot{z} = \frac{2}{1+\rho}(x(z+\rho)P_{C \to L} +y(z+\rho)P_{D \to L}-xzP_{L \to C}-yzP_{L \to D})\\
\end{array}.
\label{eqs06}
\end{equation}
\end{footnotesize}

\noindent\textbf{Remark 3: randomized bots} Let $x,y$ and $z$ denote the fractions of cooperators ($C$), defectors ($D$), and loners ($L$) in a well-mixed and infinite population such that $x+y+z=1$. We assume that bots have equal probabilities to be $C$, $D$, and $L$. Thus, the fractions of bots with action $C$, with action $D$, and with action $L$ are all $\rho/3$. The total population density is $1+\rho$. Therefore, the expected payoffs of each of the actors are as follows:
\begin{equation}
\begin{array}{l}
\Pi_{C}  = \frac{x+\frac{\rho}{3}-(y+\frac{\rho}{3})r+(z+\frac{\rho}{3})\sigma}{1+\rho} \\
\Pi_D  = \frac{(1+r)(x+\frac{\rho}{3})+(z+\frac{\rho}{3})\sigma}{1+\rho} \\
\Pi_{L}  =\sigma\\
\end{array}.
\label{eqs7}
\end{equation}
We adopt the pairwise comparison rule, where the imitation probability depends on the payoff difference between two randomly selected players. If the randomly selected players choose the same action, nothing happens. Otherwise, the probability that action $i$ replaces action $j$ is:
\begin{equation}
\begin{array}{l}
P_{j \to i}  = 1/(1+e^{(\Pi_{j}-\Pi_{i})/\kappa}) \\
\end{array}.
\label{eqs8}
\end{equation}
Where $i\not=j \in \{C,D,L\}$, and $\kappa^{-1}$ is the imitation strength such that $\kappa^{-1}>0$. The evolutionary dynamics for a well-mixed and infinite population under the imitation rule are represented by:
\begin{equation}
\begin{array}{l}
    \dot{x} = \frac{2}{1+\rho}((x+\frac{\rho}{3})yP_{D \to C} + (x+\frac{\rho}{3})zP_{L \to C}-x(y+\frac{\rho}{3})P_{C \to D}\\-x(z+\frac{\rho}{3})P_{C \to L})\\
   \dot{y} = \frac{2}{1+\rho}(x(y+\frac{\rho}{3})P_{C \to D} + (y+\frac{\rho}{3})zP_{L \to D}
-(x+\frac{\rho}{3})yP_{D \to C}\\-y(z+\frac{\rho}{3})P_{D \to L})\\
    \dot{z} = \frac{2}{1+\rho}(x(z+\frac{\rho}{3})P_{C \to L} +y(z+\frac{\rho}{3})P_{D \to L}
-(x+\frac{\rho}{3})zP_{L \to C}\\-(y+\frac{\rho}{3})zP_{L \to D})\\
\end{array}.
\label{eqs9}
\end{equation}

\bibliographystyle{elsarticle-num}
\bibliography{biblio}
\end{document}